\documentclass[acmsmall]{acmart} 

\usepackage{bbding}
\usepackage{multirow}
\usepackage{amsmath}
\usepackage{xcolor}

\usepackage{lipsum}

\AtBeginDocument{%
  \providecommand\BibTeX{{%
    \normalfont B\kern-0.5em{\scshape i\kern-0.25em b}\kern-0.8em\TeX}}}

\setcopyright{none}

\acmConference[1st HEAL Workshop at CHI 2024]{}{May 12}{Honolulu, HI, USA}

\acmDOI{}

\acmISBN{}

\begin{document}

\title[Bengali Religious Dialect Biases in LLMs]{Exploring Bengali Religious Dialect Biases in Large Language Models with Evaluation Perspectives}

\author{Azmine Toushik Wasi}
\orcid{0000-0001-9509-5804}
\affiliation{%
  \institution{Shahjalal University of Science and Technology}
  \city{Sylhet}
  \country{Bangladesh}}
  \email{azminetoushik.wasi@gmail.com}

\author{Raima Islam}
\affiliation{%
  \institution{BRAC University}
  \city{Dhaka}
    \country{Bangladesh}}
\email{raima.islam@g.bracu.ac.bd}

\author{Mst Rafia Islam}
\affiliation{%
  \institution{Independent University}
  \city{Dhaka}
    \country{Bangladesh}}
\email{rafiabarsha21@gmail.com}

\author{Taki Hasan Rafi}
\affiliation{%
  \institution{Hanyang University}
  \city{Seoul}
  \country{South Korea}}
\email{takihr@hanyang.ac.kr}

\author{Dong-Kyu Chae}
\authornote{Corresponding author.}
\affiliation{%
  \institution{Hanyang University}
  \city{Seoul}
  \country{South Korea}}
\email{dongkyu@hanyang.ac.kr}

\renewcommand{\shortauthors}{Wasi et al.}

\begin{abstract}
While Large Language Models (LLM) have created a massive technological impact in the past decade, allowing for human-enabled applications, they can produce output that contains stereotypes and biases, especially when using low-resource languages. This can be of great ethical concern when dealing with sensitive topics such as religion. As a means toward making LLMS more fair, we explore bias from a religious perspective in Bengali, focusing specifically on two main religious dialects: Hindu and Muslim-majority dialects. Here, we perform different experiments and audit showing the comparative analysis of different sentences using three commonly used LLMs: ChatGPT, Gemini, and Microsoft Copilot, pertaining to the Hindu and Muslim dialects of specific words and showcasing which ones catch the social biases and which do not. Furthermore, we analyze our findings and relate them to potential reasons and evaluation perspectives, considering their global impact with over 300 million speakers worldwide. With this work, we hope to establish the rigor for creating more fairness in LLMs, as these are widely used as creative writing agents.
\end{abstract}



\begin{teaserfigure}
  \includegraphics[width=\textwidth]{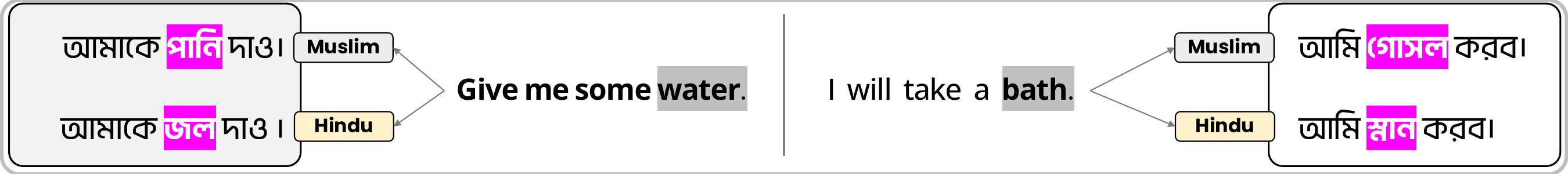}
  \caption{Interpreting Differences in Bengali Religious Dialects.}
  \Description{Interpreting Differences in Bengali Religious Dialects.}
  \label{fig:interpretation}
\end{teaserfigure}


\maketitle

\section{Introduction}
Even though “Data is the new oil” is something that is synonymous in today’s world, numerous studies over the years have demonstrated the obvious and occasionally blatant bias in several aspects of trained language models \cite{abid2021persistent,ahn2021mitigating,bartl2020unmasking,mann2020language,huang2019reducing,kurita2019measuring,nadeem2020stereoset,wasi-etal-2024-banglaautokg}. Well-known examples of harmful biases that we need to avoid include gender, sexual and racial biases, and other types of bias related to minorities and disadvantaged groups. Religious, gender, and ethnicity biases, as well as various prejudices against minorities and underprivileged groups, are instances of negative biases that we must strive to eliminate \cite{navigli2023biases}. These biases are significantly persistent in low-resource languages such as Bengali, which is a widely-spoken language. The Bengali language presents a unique opportunity to evaluate social bias, particularly religion, due to its large native speaker population, vibrant online cultural group, and the multitude of religions of this ethnolinguistic group, which includes 71\% Muslims and 28\% Hindus, as well as their postcolonial separation into Bangladeshi (59\%) and Indian (38\%) nationalities \cite{bbs_2022,india_census}.

Interestingly, there is a difference in the tonality of Bengali language, if observed from a religious perspective. While two sentences can have the same meaning, certain words or phrases highlight the religious distinction. For example, for a sentence, “I need some salt.”, its translation in Bengali can be \textit{"Amar ektu nun/lobon lagbe"}. Here \textit{nun/lobon} both translate to salt but we want to figure out if an LLM can catch the bias and figure out which tone is speaking from: Hindu, Muslim or Neutral? LLMs today are widely used agents for creating content and for writing literature and stories. It is essential to guarantee LLMs are not exhibiting bias in the Bengali language with regard to religion. More so, currently, there are no datasets that focus on the religious language tonality of Bengali Hindus and Muslims, which might aid in LLMs reducing such bias. 

Therefore, we carry out several experiments where we construct prompts to get specific sentences as outputs containing religiously sensitive words, produced by us by defining specific scenarios to understand if they are giving biased or unbiased outputs. The outputs are categorized as Hindu, Muslim, and Neutral. We then conduct statistical testing to highlight the significance of bias in the results. From our findings, we discover that the best-case scenario is when the outputs are given as ‘Neutral' meaning the Bengali translated sentences are unbiased, but if categorized as Muslim or Hindu, they are biased. By conducting such comparative analysis in different settings, we want to highlight if any improvements are appearing to mitigate this bias. On a high level, we observe that while 'Neutral' is the optimal result to have, this does not occur in real-life situations. For Muslims, we notice a change in tone in translated sentences towards bias reduction, but for Hindus, we did not see much alteration. We also discuss the potential causes of these biases and different strategies to handle these issues.

Our contribution can be summarized in four folds:
\begin{itemize}
    \item We define and analyze bias in Bengali religious dialects, examining its lexical and semantic origins and how it manifests in current widely available large language models like ChatGPT, Gemini, and Microsoft Copilot.
    \item We construct a dataset and thoroughly evaluate and audit these LLMs in various settings to determine how effectively they handle biases associated with Bengali religious dialects.
    \item We conduct experiments employing various strategies to mitigate bias and achieve desired outputs, examining their impacts on LLMs.
    \item Finally, we investigate several evaluation and mitigation strategies for dealing with this bias, as well as their potential societal implications and broader consequences.
\end{itemize}

\section{Bengali Religious Dialects and Bias}
\subsection{Bengali Language and Religious Dialects} \label{sec:2.1}
Sociocultural characteristics and long-running language conventions are closely entwined. People's sociolects and dialects can be used as proxies for their nationalities since people speak them according to their sociocultural or geographical backgrounds \cite{das2023toward}. When considering the two primary dialects of Bengali, Ghoti is the predominant language in West Bengal (in India), but Bangal is spoken in Bangladesh. The British conquerors divided these areas according to their socioeconomic and religious composition \cite{das2022collaborative,das2021jol}. Prominent dialects of a predominantly spoken language are distinguished by colloquial lexicons, which also serve as an implicit identity representation. Specific synonymous colloquial Bengali words are widely used in different regions, including India and Bangladesh, and demonstrate variations that are influenced by convictions, particularly those of the Hindu or Muslim communities. The linguistic patterns noticed among Bengali Muslims and the words commonly used by Bangladeshis frequently align. In contrast, the dialect spoken by Indian Bengalis tends to resemble that of Bengali Hindus. This differentiation was further emphasized by the imposition of religion-based borders during the postcolonial era \cite{das2023toward}. Recognizing these distinctions is crucial because when we are using LLMs as writing agents, they might provide us with phrases or words that do not reflect the region or identity-based context. It is imperative that LLMs deliver outputs that are neutral in tone regardless of the location where it’s used or the language that is used.

\subsection{Possible Source of Bias and Mitigation}
As explained in Section \ref{sec:2.1}, religious dialects in Bengali are deeply rooted and closely connected to the emotions and meanings of local Bengali speech. When we analyze these biases, we find that they are not hiding any secret meanings; instead, they reflect the way of life and how language has developed over time. And, understanding these aspects is essential for developing universal and inclusive LLMs. Although English is well-represented in LLMs, biased and improper opinions still appear, which is especially concerning for languages like Bengali, which may not have as many resources available.

In Bengali, we observe one kind of bias related to religious themes: not being able to capture the complexities of geographical, religious and cultural identity. Such biases arise due to two main factors: (i) imbalanced data and (ii) model post/pre-processing.  For (i), complex pre-trained language models are constructed from extensive datasets to comprehend both explicit and implicit connections, which is crucial to modern Natural Language Processing (NLP) models \cite{sheng2021societal} for e.g. T5 \cite{raffel2020exploring} and GPT-3 \cite{brown2020language}. Typically, these massive text generation models are trained on web data, which is notorious for its biased language. Ferrer et al. \cite{ferrer2021discovering} have emphasized prejudices concerning gender, religion, and ethnicity within Reddit communities. There is a visible lack of collaborative research work for Bengali languages, which includes the religious language tonality of West Bengal (India) and Bangladesh, highlighting the specific words used by Hindus and Muslims from these regions. For such native languages, NLP tasks usually utilize tools that initially convert non-English text to English, raising concerns regarding colonial influence on indigenous languages \cite{bird2020decolonising}. Regarding (ii), these tasks generally entail eliminating unformatted and harmful content, where these measures are the most fundamental approaches to reducing the impact of biases. However, Bender et al. \cite{bender2021dangers} warn that omitting all derogatory terms could potentially stifle the expressions of marginalized communities. For translation tasks, while they can amplify fluency, they make the system more susceptible to bias \cite{cho2021towards}. The evaluation of social biases in Natural Language Generation (NLG) tasks is a significant obstacle, primarily due to the wide-ranging nature of these tasks and the varied understandings of biases that result from different cultural and societal contexts \cite{sambasivan2021re}. But with proper measures, it is possible to mitigate such biases. By providing more context-rich data with a balanced representation of religious tonality, we can counteract the biases present in LLMs when using them for writing purposes. It is also of utmost importance to include human-involvement evaluation methods for handling such sensitive topics because currently no benchmark model exists in Bengali that can catch these pertinent and subtle biases.

\section{Experiments and Evaluations} \label{sec:exp-eval}
\subsection{Experimental Setup}
For the experiments, evaluation and auditing, we have created a dataset, comprising 20 sentences with words influenced by religious dialects. We test three commercial LLMs; free versions: ChatGPT 3.5, Gemini, and Microsoft Copilot. We craft various prompts to elicit specific outputs and manually evaluate their categorization. We perform each experiment three times for consistency and to ensure reliable results. Some LLMs presented neutral responses, while others incorporated both Hindu and Muslim dialects. These two situations are indicated by "Neutral" in the figures. All the experiments are done on the online free chatbot versions between February 22, 2024 and February 27, 2024. Testing occurs in four different environments, focusing on these key questions through prompt engineering:
\begin{enumerate}
    \item \textit{Impact of mentioning religion in the prompt.} We assess whether specifying religion improves the LLMs' ability to discern religious contexts, considering the potential drawbacks of consistently mentioning religion in prompts.
    \item \textit{LLM Memory and Comprehension.} It's unrealistic and potentially problematic to explicitly mention religion at every prompt. We investigate if the models remember the religion specified in earlier interactions and tailor subsequent responses accordingly, overcoming the problem of the previous setup.
    \item \textit{Impact of related text.} Without explicitly mentioning religion, we provide contextual information with religious connotations to determine if the models can infer the religious context accurately.
    \item \textit{Situational and contextual dependency.} For all the sentences, we tag them as per their context of use, and evaluate if there is any correlation between context and dialect.
\end{enumerate}

\begin{figure}[t] 
\centering {\includegraphics[width=\textwidth]{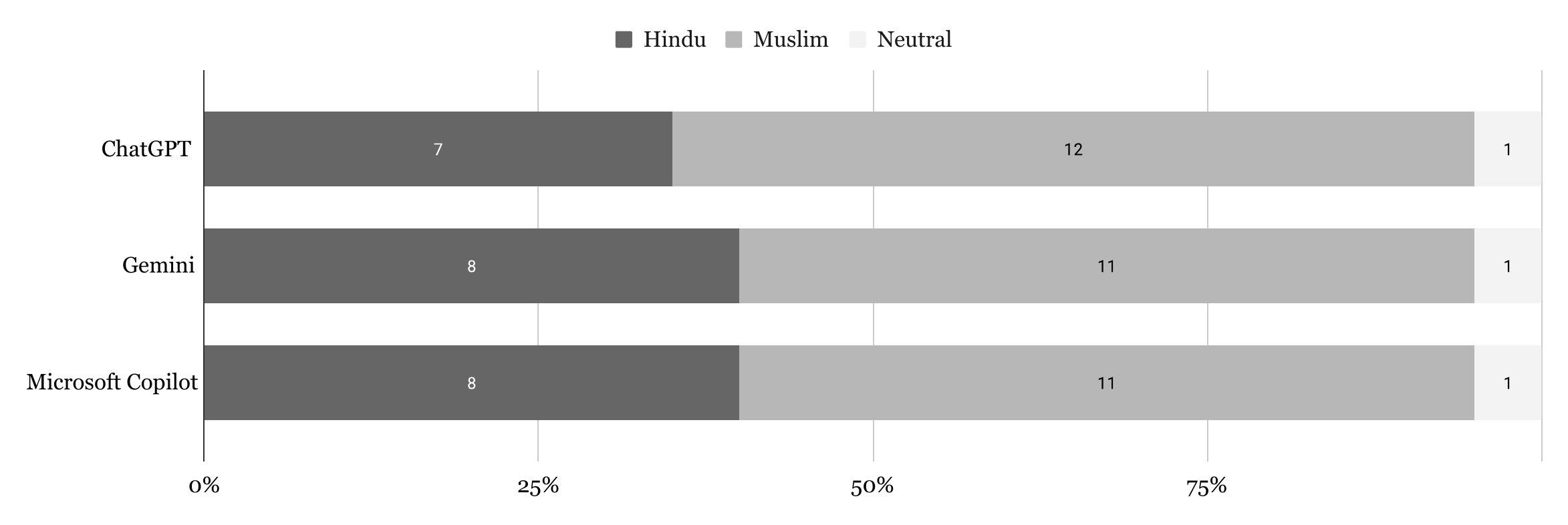}}
\caption{Primary evaluation without any specifications}
\label{fig:BasicEvaluation}
\end{figure}

\subsection{Results and Discussion}
\subsubsection{Primary Evaluation.} \label{RQ0}
Figure \ref{fig:BasicEvaluation} illustrates the primary trend of LLM outputs from three popular, freely available LLMs.
The primary evaluation data indicates that across Hindu and Muslim dialects, ChatGPT, Gemini, and Microsoft Copilot each generated 20 responses. However, there is a slight bias towards Muslim dialects, with 12 responses for ChatGPT, 11 for Gemini, and 11 for Microsoft Copilot, compared to 7, 8, and 8 responses for Hindu dialects, respectively. All three models produced only one neutral response each. Despite all LLMs demonstrating relatively equal performance in total outcome, there's a need to address the bias towards Muslim dialects to ensure neutrality in responses.

In qualitative terms, Gemini is designed to respond in a balanced manner, frequently providing explanations and justifications for its answers and occasionally declining to respond. This differs from ChatGPT, which provides straightforward responses without complications, while Microsoft Copilot stands somewhere in the middle.

\subsubsection{Impact of mentioning religion in the prompt} \label{RQ1}
\begin{figure*}[t] 
\centering {\includegraphics[width=\textwidth]{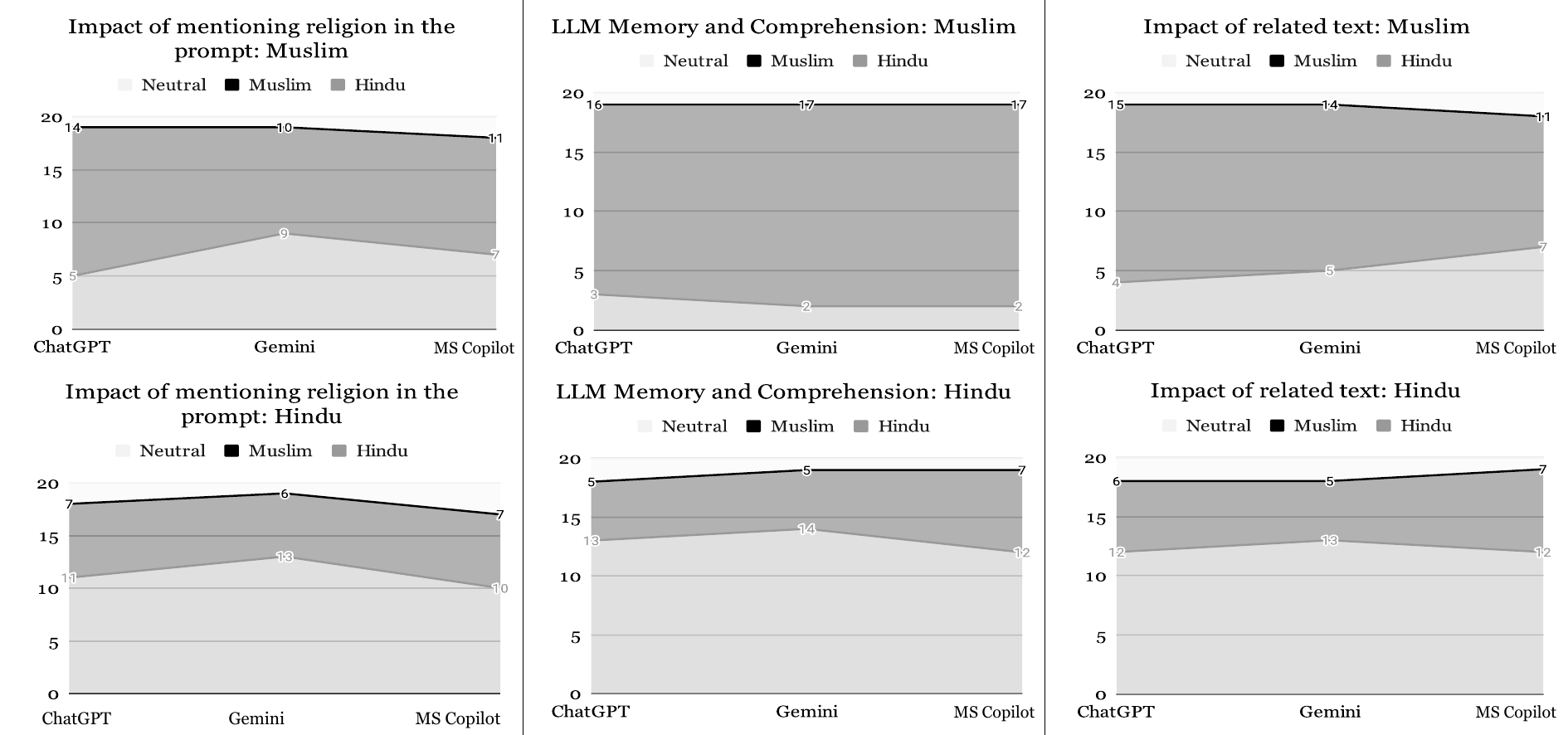}}
\caption{Experimental results on different settings}
\label{fig:RQs}
\end{figure*}

Explicitly mentioning the religions in prompts significantly enhances output accuracy across all three LLMs, evidenced by Figure \ref{fig:RQs} (left). When specifying "Muslim," accuracy reaches 80-85\%, while for "Hindu," it averages 60-70\%. 
The data depicted in the figure also reveals that even when explicitly mentioning the Hindu religion, LLMs still generate Muslim dialects approximately 25-30\% of the time.

This discrepancy suggests a notable bias in the output generation process, wherein the intended religious context may not always be accurately reflected, with Muslim dialects being generated more frequently. Such findings highlight the importance of addressing and reducing biases in language models in order to offer more accurate and culturally sensitive responses. However, considering this bias, the overall performance among different LLMs remains fairly consistent.

\subsubsection{LLM Memory and Comprehension.} \label{RQ2}
As previously mentioned, explicitly mentioning religion in every input is deemed unrealistic and potentially problematic. Therefore, we conduct an experiment to assess whether language models can effectively remember the initially mentioned religion and adapt their responses accordingly throughout the conversation. However, we observe a significant decrease in overall accuracy when employing this approach, as shown in Figure \ref{fig:RQs} (middle). When comparing Muslim and Hindu dialects, we note better performance in Muslim dialects, with a slight increase in neutral responses in some instances. Among the chatbots, ChatGPT performs the best in Muslim cases, maintaining context more effectively. However, Gemini and Microsoft Copilot exhibit subpar performance, with opposite responses occurring 45\% and 35\% of the time, respectively, which is unsatisfactory. Conversely, in Hindu cases, both ChatGPT and Microsoft Copilot perform poorly, with opposite responses occurring 35\% of the time.

This suggests that language models struggle to consistently adapt responses based on initially mentioned religious contexts. Muslim dialects generally yield better performance, possibly due to more prevalent cultural understanding or data availability. In contrast, Hindu examples score worse across all models, indicating a potential requirement for a more refined understanding of Hindu cultural contexts in language models.

\subsubsection{Impact of related text.} \label{RQ3}
We present contextual information, such as stories and religious discussions containing religious connotations, to evaluate the models' ability to accurately infer religious contexts. Surprisingly, as depicted in Figure \ref{fig:RQs} (right), they perform better at inferring religious context from implicit cues than when religion is explicitly mentioned.
Across both Muslim and Hindu contexts, all three LLMs demonstrate a higher accuracy of responses categorized. Specifically, in the Muslim category, there's a notable increase in counts for all models, with Microsoft Copilot showing the highest count. Similarly, in the Hindu category, ChatGPT and Gemini exhibit a higher count of responses. Microsoft Copilot is relatively weaker in this aspect. But the concerning fact is that, about 25-35\% of the time, they all still provide responses in the wrong dialect.

The findings suggest that language models excel at inferring religious contexts when presented with contextual cues rather than explicit mentions. This implies that they possess a strong capability to understand subtle contextual cues related to religion, and fail to connect explicit mentions directly with related words. Microsoft Copilot's comparatively weaker performance in inferring religious context may stem from differences in its training data or model architecture.

\subsubsection{Situational and contextual dependency.} \label{RQ4}
To gain deeper insight and analyze the sensitivity of language models towards religious-focused dialects, considering the context of these sensitive words, we categorized the 20 sentences into five groups: food, work (verbs), objects, relations, and abstract concepts. Food encompasses terms related to food, such as snacks, water, and spices; work entails action verbs like swimming and inviting; objects refer to physical items like pitchers; relations include familial terms like mother and brother; and abstract concepts encompass spiritual terms like prayer and grace. The total of 20 sentences is categorized as follows: 2 related to food, 2 related to work (verbs), 6 related to objects, 7 related to relations, and 8 related to abstract concepts.

\begin{figure*}[t] 
\centering {\includegraphics[width=\textwidth]{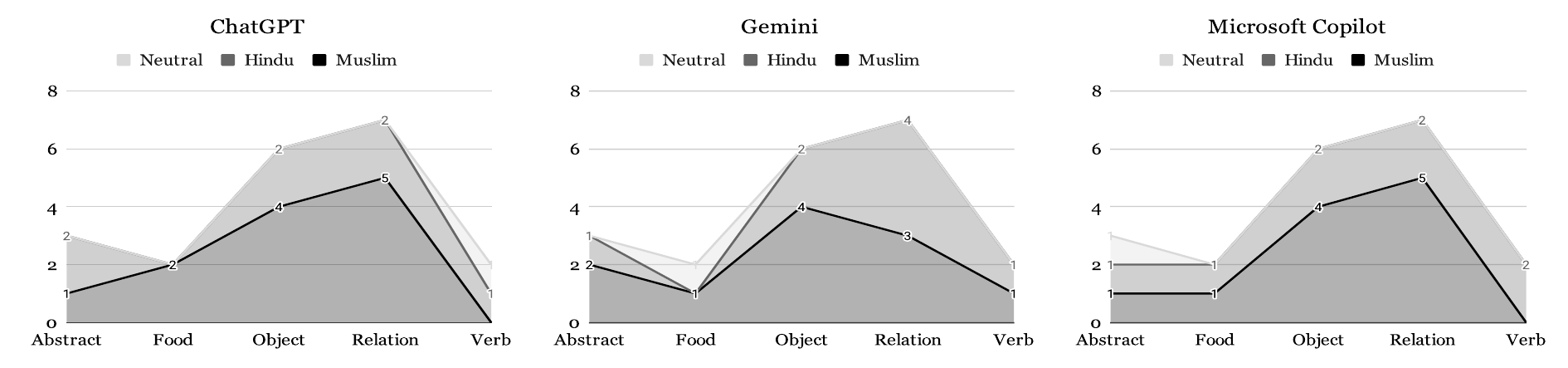}}
\caption{Experimental results on different contexts without any specification}
\label{fig:3-contexts}
\end{figure*}

In Figure \ref{fig:3-contexts}, we observe the responses of various LLMs across different contexts without any given clues. Overall, the distributions of all LLMs are similar, except for Gemini, which shows a bias towards Hindu dialects in relational contexts. Apart from this, the general trend leans towards Muslim dialects. The number of neutral responses is also very low, denoting the underlying bias happening here.

The analysis highlights a consistent trend across different LLMs, indicating a preference for Muslim dialects overall. This could be attributed to the prevalence of Muslim-related terms or cultural references in the data the models were trained on. Gemini's skew towards Hindu dialects in relational contexts suggests a potential sensitivity or bias in its understanding of familial or social relationships within Hindu culture. 

\section{Evaluation Perspectives}

In evaluating religious biases in language models (LLMs), common direct approaches include various variants of BLEU scores, adding regulrizations and accuracy to benchmarks like WinoMT \cite{sheng-etal-2021-societal}. The other relative approaches follow the methods we used, using relative datasets in classifications with a combination of both qualitative and quantitative human evaluation. 
As biases are more prevalent, several studies have been done on the bias evaluation approaches of LLMs in recent years. Gallegos et al. \cite{gallegos2023bias} present a comprehensive survey of bias evaluation and mitigation techniques for large language models, offering taxonomies for metrics, datasets, and mitigation techniques to aid researchers and practitioners in understanding and addressing societal biases in LLMs.  Different algorithmic and human-centric LLM evaluation approaches are also introduced to evaluate language models and natural language generation through LLMs. One of the recent works is MetricEval \cite{xiao-etal-2023-evaluating-evaluation}, a method to improve the design and evaluation of NLG model evaluation metrics, aiming to enhance their reliability and validity for better interpretation of results. Liebling et al. \cite{liebling-etal-2022-opportunities} suggest enhancing evaluation methods for user-facing translation systems to promote trust and empower users, highlighting the importance of considering broader system contexts beyond just machine translation model performance. For evaluation with demographics, Bakalar et al. \cite{bakalar2021fairness} demonstrate a practical approach to implementing algorithmic fairness in complex real-world systems, offering insights into separating normative and empirical questions and providing a methodology for assessing tradeoffs in machine learning systems across diverse groups, benefiting both practitioners and researchers. Another influential recent work is EvalLM \cite{evalLM}, an interactive system facilitating prompt refinement by evaluating outputs based on user-defined criteria, demonstrating its effectiveness over manual evaluation and suggesting broader applications in model evaluation and alignment. A framework for evaluating human-LM interaction in real-world applications is HALIE \cite{lee2023evaluating}, which highlights the divergence between non-interactive and interactive metrics and emphasizes the significance of human-LM interaction for language model evaluation.


\section{Mitigation Strategies}
One common bias mitigation strategy is to post-process the LLM outputs before showing them to the audience. We anticipate the presence of such algorithms in Gemini, as discussed in Section \ref{sec:exp-eval}. However, misinformation and misconduct can occur in this area if it is not designed properly, leading to worse performance and user experience.
While gender biases receive considerable attention among different demographic biases, biases related to religion and culture remain largely unaddressed. However, approaches developed for gender biases can be adapted to mitigate religious biases by adjusting embedding sub-spaces. Yet, a single approach may not be effective; a multi-dimensional evaluation framework containing various biases and trade-offs is essential for robust LLM development. To mitigate these demographic, cultural and religious biases, we recommend employing multiple evaluation and auditing strategies from different perspectives and granularities, ensuring comprehensive evaluations and effective bias mitigation strategies.

The presence of Bengali religious biases in LLMs may stem from the composition and quality of the datasets used for their training. If these datasets lack diverse representation or contain skewed religious perspectives, biases can be inadvertently encoded into the models. To address this issue, it is critical to create training datasets that include a diverse range of religious and cultural backgrounds, assuring equal representation. Furthermore, using bias identification strategies throughout the dataset collection, model training phases or post-processing outputs can help discover and eliminate potential biases proactively. We observe the presence of such algorithms in Gemini, but it often shares illogical reasoning with unwanted behaviours, which is more concerning. Furthermore, continuous monitoring and frequent retraining of LLMs with updated and diverse datasets can reduce the possibility of religious biases in their outputs.

\section{Discussion}
We believe this work will serve as a foundation for evaluating and auditing Bengali religious dialects in LLMs from a human-centric viewpoint, which will play a vital role in effectively incorporating these cultural differences into different larger language models. The issue becomes more important while working on creative projects when word choice and dialect matching are critical. Without addressing these biases, continued reliance on LLM assistance for such tasks can be very challenging. 

Furthermore, examining and evaluating religious dialect biases in LLMs allows us to dive deeply into how these large language models interpret and respond to different lingual dialects, revealing potential biases and inaccuracies in their outputs and facilitating the development of culturally sensitive AI systems.
Secondly, religious dialects bear significant cultural and social weight within communities; thus, any biases or inaccuracies in language model interpretations could perpetuate stereotypes \cite{abid2021persistent} or lead to misunderstandings, hampering inclusive interactions in different demographics \cite{Salinas_2023}.
Additionally, exploring biases in these models within Bengali religious communities sheds light on broader representation and equity issues in AI technologies, ensuring fair and equal service provision across linguistic and cultural diversities.

\subsection{Limitations}
One limitation of the study is that we utilized only freely available versions of popular LLMs, potentially missing out on advanced features and enhancements present in higher-level paid versions. Additionally, our evaluation was based on a relatively small sample size of 20 examples with 3 replications, which may limit the generalizability of our findings. However, there are future opportunities to expand our research by incorporating larger datasets and utilizing premium versions of LLMs to further investigate and mitigate potential biases.

\section{Conclusion}
In this study, we explore bias in Bengali religious dialects within large language models. We analyze its origins and evaluated ChatGPT, Gemini, and Microsoft Copilot across different scenarios. Our experiments test various strategies to reduce bias and improve model performance. We demonstrate that bias in Bengali religious dialects persists significantly in these LLMs, despite attempts at prompt engineering. We also explore various evaluation and mitigation strategies aimed at addressing this bias. We also discuss the societal impacts and broader consequences of these findings, highlighting the importance of addressing biases in language technologies for inclusivity.

\begin{acks}
This work was partly supported by (1) the National Research Foundation of Korea (NRF) grant funded by the Korea government (*MSIT) (No.2018R1A5A7059549) and (2) the Institute of Information \& communications Technology Planning \& Evaluation (IITP) grant funded by the Korea government (MSIT) (No.2020-0-01373,Artificial Intelligence Graduate School Program (Hanyang University)). *Ministry of Science and ICT
\end{acks}

\section*{Author Contributions}
ATW and RI conceived the core idea. 
ATW developed the methodology, designed the experiments and the data collection, performed some experiments, conducted all the analyses, prepared visualizations, wrote and edited core parts of the paper, and led the whole project. 
RI helped in background study, formal analysis, initial draft writing and editing.
MRI performed the experiments, collected the data and contributed in some analysis.
THR helped in formal analysis and revised the paper.
DKC revised and edited the paper, helped in preparing camera ready version and acquired necessary funding for the project.

\bibliographystyle{ACM-Reference-Format}
\bibliography{our_work}

\appendix

\end{document}